\documentclass[12pt,preprint]{aastex}

\usepackage[]{natbib}

\shorttitle{Pion production in accretion disk}
\shortauthors{Meirelles et al.}

\begin{document}


\title{Pion production In The Inner Disk Around Cygnus X-1}




\author{Cesar Meirelles Filho}

\affil{Instituto de Astronomia, Geof\'{\i}sica e de Ci\^{e}ncias Atmosf\'{e}ricas \\
Universidade de S\~ao Paulo \\R. do Mat\~ao, 1226, 05508-090 S\~ao Paulo, SP, Brasil}

\email{cmeirelf@astro.iag.usp.br}

\author{Celso L. Lima, Hideaki Miyake, and Varese Timoteo}

\affil{Instituto de F\'{\i}sica \\Universidade de S\~ao Paulo \\CP 66318, 05315-970 S\~ao Paulo, SP, Brasil}

%
%
%




\begin{abstract}

Neutron production via ${}^{4}$He breakup and $p$${\left( p,\,n{\pi }%
^{+}\right) }$$p$ is considered in the innermost region of an accretion
disk surrounding a Kerr Black Hole. These reactions occur in a plasma in
Wien equilibrium, where (radiatively produced) pair production equals annihilation.
Cooling of the disk is assumed to be due to unsaturated inverse
Comptonization of external soft photons and to the energy needed to ignite
both nuclear reactions. Assuming matter composition of $90\%$ Hydrogen and $%
10\%$ He, it is shown that, close to the border of this region, neutron
production is essentially from $^{4}$He breakup. Close to the horizon, the
contribution from $p\,{\left( p,\,n{\pi }^{+}\right) }\,p$ to the neutron
production is comparable to that from the breakup. It is shown that the
viscosity generated by the collisions of the accreting matter with the
neutrons, may drive stationary accretion, for accretion rates below a
critical value. In this case, solution to the
disk equations is double-valued and for both solutions protons overnumber
the pairs. It is claimed that these solutions may mimic the states of high
and low luminosity observed in Cygnus X-1 and related sources. This would be
explained either by the coupling of thermal instability to the peculiar
behavior of the viscosity parameter $\alpha $ with the ion temperature that
may intermittently switch accretion off or by the impossibility of a perfect
tuning for both thermal and pair equilibrium in the disk, a fact that forces
the system to undergo a kind of limit cycle behavior around the upper
solution.

\end{abstract}




\section{Introduction}

Accretion disks are presently thought as the scenario for hydrodynamical
flows in close binary systems, galactic nuclei and quasars. Despite this
fast growing recognition, models which allow for some particle processes,
taking into account the detailed nature of the flow, as well as realistic
processes that may act as the source for the required viscosity, are still
lacking. Concerning particles processes, most of the theoretical work on the
field has focused on the production of positron-electron pairs. Steady state
scenarios, under the assumption of production-annihilation equilibrium, have
been first tackled by \citet{bis71}, who found that, if
particle-particle dominates pair creation, equilibrium is only possible for
lepton temperature below $20m{c}^{2}$, \citet{poz77}, \citet{stoe77}, 
and by \citet{lian79}, who considered pair production dominated by
gamma-gamma interactions, in a plasma under Wien equilibrium, finding
multivalued solutions to the disk structure. These results, however, are
applicable only to a small region of parameter space or to a subset of
important reactions. More general studies have been subsequently carried out
by many authors \citep{lig82,sve82,sve84,zdz84,tak85,lig87,whi89}, 
which contributed to a better understanding of radiative
processes in very hot astrophysical plasmas, opening the way to the
understanding of the topology of the disk equations solution, which at fixed 
$r$ in the ${\dot{M}}\,{\Sigma }$ plane forms an S-shape kink corresponding
to its multiple valued nature \citep{kus90,lian91,bjo91,bjo92,min93,kus92,kus95,kus96}. 
This multivalueness consists of three branches, two
of them being hot (one pair dominated and the other pair deficient) and the
third being cool and pair deficient. The hot pair dominated branch is very
promising to explain some especial features in some black hole candidates,
like the MeV bump of Cygnus X-1, the radio plasmoid bipolar outflow of the
jet sources, and the alleged annihilation line features \citep{lin87,lian98,
mir92,mir94,hje95,paul91}. It may happen that, under the conditions prevailing in the
innermost parts of some systems, the time needed for the matter to cross
these regions is comparable to, or even less than, the time the electrons
need to extract energy from the protons, through collisional energy
exchange. These regions will grow in the direction perpendicular to the
plane of the disk, becoming geometrically thick (swelling of the disk may
also occur for systems with Eddington or super Eddington luminosities 
\citep{abr80,pac82}, not considered in
this paper). From angular momentum transport point of view, collective
processes are efficiently operative to generate viscosity (due to the
protons), but with a deficient cooling (due to the electrons). This drives
protons and electrons out of thermal equilibrium, the ion temperature being
much greater and close to the virial one. For a disk around a maximally
(synchronously) rotating Kerr black hole the inner radius of the disk may
reach the horizon, with the ion temperature approaching the mass of the
proton. A two temperature geometrically thicker and optically thin disk
model was first developed by \citet{shap76} to explain
the hard X-ray from Cygnus X-1. These authors, however, have not considered
radial advection of energy and entropy which, in that situation, are very
important and the local approximation breaks down. Advection of energy in
optically thin disks was first considered by \citep{ichi77}, in the context
of the bimodal behavior of the X-ray spectra from Cygnus X-1, \citep{lian80}
 who considered advection in an optically thin,
two-temperature disk, and by \citep{whi90} who studied advection
of energy and pairs in optically thin disks. \citep{pac81,muc82,abr88,hon91,wal91,chen93}, 
have unraveled the basic physics
of advection in accretion flows with their studies on the properties and
structure of optically thick disks. However, only recently, advective
cooling in optically thin accretion flows has been recognized thanks to a
systematic work by \citet{nar94,nar95}, \citet{nar95a,nar95b}, \citet{abr95}, 
\citet{nar95a,nar95b}, \citet{nar96}, \citet{las96a,las96b}, and \citet{chen95}. 
Neglecting pairs, these authors have studied the topology of the
solution in the ${\dot{M}}\,{\Sigma }$ plane, showing the existence of a
maximum accretion rate, above which no steady state solution is allowed, and
that, below it, there are two optically thin solutions, namely the radiative
cooling dominated and the advective cooling dominated states. A general
description of flows in accretion disks, taking advection into account and
neglecting pairs, was given by \citet{chen95}. They have shown that, at a
given radius, there exist exactly four physically distinct types of
accretion disks. Two of these correspond to values of ${\alpha }$, the
viscosity parameter, smaller than a critical value, and the other two, to
values of ${\alpha }$ greater than this critical value. Inclusion of pairs
have been considered by \citet{bjo96}, \citet{kus96}, \citet{esin96,esin97}, 
who have shown that pairs do
not modify the topology of the solution, besides being negligible for ${%
\alpha }<1$. It should be remarked that, though the advective model
constitutes an improvement as far as the standard Shakura and Sunyaev disk
model is considered, till the moment, no global solution exists for the disk
\citep{bjo96}. At this point, we must realize the importance
of the viscosity parameter and, yet, we don't know the physical process
that may be the source for such a viscosity. Since the seminal paper by
\citet{shak73}, a lot of mechanisms have been proposed to account
for viscosity: shear turbulence generated by the Keplerian rotation of the
disk \citep{shak73,zel81,dub90,zahn91}, turbulence driven by convection 
\citep{lin80,tay80,bis77,shak77,ryu92,cab92,sto96,mei91a,mei93,mei97}, 
tangled magnetic fields sheared by
differential rotation \citep{lyn69,shak73,ear75,ichi77,cor81}, 
angular momentum transport by waves \citep{pap84a,pap84b,vis89,vis90},
Velikov-Chandrasekhar magnetic instability \citep{bal91,bal92,bal96,vis92},
ion viscosity and neutron viscosity \citep{gue90,mei93}, radiative viscosity 
\citep{loe91}. None of these processes, 
however, is immune to criticism: concerning shear turbulence, disks satisfy
Rayleigh's criterion for stability; convection may transport angular
momentum inwards rather than outwards; magnetic field may be removed by
magnetic buoyance and ohmic dissipation; concerning wave turbulence, it
seems that the most stable mode have low wavenumbers, giving rise to
structures of the order of the size of the system, being very sensitive to
the Coriolis force which hinders the appearance of smaller structures; the
Velikov-Chandrasekhar instability is highly dependent on the radial
structure of the azimuthal component of the magnetic field, and can only
occur if the Alfv\'{e}n velocity is of the order of the Keplerian one, or if
the typical scale of the magnetic field is smaller than the Keplerian one;
as far as ion viscosity is concerned, besides needing high temperatures 
(the system needs another process to achieve these high temperatures), any
small magnetic field present in the flow will decrease the strength of the
viscosity by orders of magnitude; neutron viscosity needs high temperature
to ignite nuclear reactions capable{\em \ }to produce neutrons, needing,
therefore, some other process{\em \ }to heat up particles till these high
temperatures; Keplerian thin disks cannot be supported by radiative
viscosity, because the required energy density of the photon gas should be
one or two orders of magnitude larger.

In view of all these uncertainties concerning the hydrodynamics of the flow,
as well as the angular momentum transport in accretion disks, we would like
to consider neutron viscosity, assuming neutron production through Helium
break up and through pion production due to proton-proton collisions. The
threshold for Helium breakup is about 20 MeV and about 290 MeV for the
production of pions (and neutrons) in the proton-proton collision. Close to
the horizon, the energy in the rest frame of two colliding protons is about
2 GeV, which means that, at least theoretically, these reactions are
energetically viable. The real difficulty is to find conditions in
parameter space, if they exist, such that the plasma is no longer ruled by
electromagnetic interactions alone and nuclear interactions start playing a
role. Effectiveness of strong interactions depends not only on the value of
ion temperature itself but also on the electron temperature. This is so
because electromagnetic interactions are restricted to the Debye sphere and
the number of electrons within it decreases with electron temperature. High
electron temperature will favor, not only the production of
electron-positron pairs, but also the possibility of nuclear reactions. One
has to realize we are here dealing with a very intricate situation highly
dependent on the viscosity. It is well known, that for a given density,
electron and proton temperatures, drift time decreases with increasing
viscosity parameter while $t_{ep}$, the electron-proton collision time, has
inverse behavior. However, physical variables in the disk are very sensitive
functions of the viscosity parameter and we have to make a self-consistent
calculation to find a region in parameter space where the nuclear reactions 
effects are maximized.

Account for nuclear reactions in the accretion disks may have interesting
observational consequences such as the lines spectral features of
Hydrogen-Helium plasmas at high temperatures and $\gamma $-ray lines
production. From the theoretical point of view, Helium breakup and pion
production reactions produce neutrons whose collisions with the accreting
matter may be a source for the viscosity in the disk, which, in turn, will
depend upon electron temperature, proton temperature and accretion rate.
This may have interesting consequences as far as the topology of the
solution is concerned. This is a problem we want to tackle in a future work.
However, in the present paper, it should be stressed, our main concern is to
find out if neutron production through proton-proton collisions is viable in
the innermost region in accretion disks, and if its collisions with the
accreting matter can drive accretion on. We will, therefore, be interested
only in order of magnitude estimates. Keeping this on mind, we will make
some reasonable approximations, mainly on the radiative transport, pair
production and on the hydrodynamics, which, at the right moment, will be
justified.

A potentially promising application of the afore mentioned reactions is the
modelling of some X-ray systems, like Cygnus X-1, that exhibit multimodal
behavior. In a previous paper \citep{mei93}, it was shown that although
these reactions, in steady state accretion, can not supply the disk with
enough neutrons to make their collisions with the accreting matter the main
source of viscosity, they may have strong implications as far as the
temporal behavior of the disks is concerned. We have considered, however,
only the production of neutrons through the ${}^{4}$He breakup reaction 
$$
{}^{4}He+28.296 MeV \rightarrow 2p+2n ,\eqno(1) 
$$%
for ion temperatures greater than $3$ MeV. Assuming that a steady accretion
can be achieved with the drift time equal to the nuclear reaction time, we
were able to show the existence of a critical accretion rate, below which
there is no steady state accretion onto the hole. Above the critical
accretion rate it is possible, under special circumstances, for the disk to
choose between the two states of steady accretion.

This kind of procedure has to be criticized on the grounds that equality
between drift and nuclear reaction times is a very stringent constraint and,
at the temperature range considered, electron-positron pair production
should be taken into account. Besides, as we approach the hole, ion
temperature can be much greater and even exceed the threshold temperature
for the reaction 
$$
p+p+290 MeV\rightarrow p+n+\pi \,\,.\eqno(2) 
$$

We claim that accounting of both reactions in the inner parts of an
accretion disk together with allowance for thermal instability may explain
the transitions observed in Cygnus X-1, between the states of high and low
luminosity.

\section{Observational aspects: a brief review}

There is a widely accepted suspicion that the unseen compact object in the
binary system Cygnus X-1 is a black hole \citep{lian98}. Support for this 
suspicion is found on detailed analyses carried on radial velocity
measurements as well as on recent analyses based on
spectrum synthesis \citep{her95} both of which give a mass of about $10\,M_{0}$,
comfortably larger than the $3\,M_{0}$ upper limit of a neutron star mass.
The luminosity is sub-Eddington, but greater than $0.01\,L_{Edd}$ \citep{lian84}. 
Cygnus X-1, being one of the most studied sources in the sky,
especially in the hard X-rays, some of its properties (characteristic hard
X-ray spectrum, episodic emergence of an ultra soft component,
anticorrelated soft and hard X-ray transitions, chaotic variability down to
milliseconds, persistent gamma-ray tail above an MeV, episodic gamma-ray
bumps at a few hundred keV-MeV, persistent radio emission and radio flaring
correlated with X-ray transitions and Low-Frequency Quasi Periodic
Oscillations (QPO)) have been accepted as evidence for black hole candidacy
\citep{lian98}. Observations usually find the system in one of the two states:
the hard state and the soft state \citep{oda74,lian84,tana95}. 
Most of the time, Cygnus X-1 is found in the hard state,
where its soft X-ray (2-10 KeV) is relatively low and the hard X-ray
spectrum is hard. After being in this state for a few years, the system
undergoes a transition to the soft state, increasing its soft X-ray flux by
a factor of about 10, softening its X-ray spectrum. It remains in the soft
state from weeks to months before returning to the hard state. The
transition between the two states lasts from less than a day to more than a
week \citep{wen01}. In the hard state, Cygnus X-1 has a power law
spectrum characterized by a photon index $\approx 1.4$. During the hard
state the soft X-ray spectrum below 10 KeV is often a simple continuation of
the hard X-ray power law, with increasing flattening below 3 KeV. In that
case, the entire X-ray continuum is likely produced by a single hot
component. However, during some episodes of Cygnus X-1 hard state, the best
fit model still requires a small black-body component of temperature a few
tenths of KeV on the top of the power law, presuming the existence of
another region of lesser temperature \citep{lian98}. This is highly suggestive
of a soft photon source, as expected in the inverse Compton model. In
addition there is often evidence of the FeK-edge absorption at $\approx $ 7
KeV \citep{lian98,esin98}. When Cygnus X-1 is in the soft state,
its spectrum switches to one that is dominated by an ultrasoft component,
with a temperature $\approx 0.3-0.4$ KeV. The spectrum above 10 KeV becomes
much softer than in the hard state, with a variable photon index $\approx
1.9-2.5$ \citep{gie97,cui97,dot97}. It
exhibits, also, pair annihilation features in the region $500<E<1000$ KeV
\citep{lian93}.




\section{The model}
The region we shall be concerned extends away from $R_i$, inner radius of
the disk, to approximately $10$ to $15\,R_i$, where $R_i\approx .5\,R_S$, $
R_S$ being the gravitational radius.

We shall assume conditions such that the ion temperature, $T_{i}$, is much
greater than $T_{e}$, the electron temperature, with pressure given by the
ions. In the outermost parts of the disk, matter composition is Hydrogen
(nine tenth) and Helium (one tenth). In its way down to the hole, He
depletion starts and the neutrons produced gradually contributes to the
viscosity. As the incoming matter comes closer to the hole, charge exchange
reactions start to contribute to the neutron production. Concomitant to
these reactions, electron-positron pairs are radioactively produced. We
shall assume production-annihilation equilibrium. We shall also assume an
external source of soft photons that are continuously impinging upon the
disk. Part of these soft photons succeed being upscattered in energy,
reaching the Wien peak where they interact and produce pairs. The remaining
are upscattered, but leave the disk before reaching the Wien peak. Both
nuclear reactions we are considering are endothermic, so they lower the
plasma temperature by taking the energy they need to ignite. For the Helium
breakup, the number of the resulting particles is greater than the number of
the interacting particles. These particles have to be thermalized and so act
as an additional source of cooling. The pion produced by the charge exchange
reaction rapidly decay, producing a positron that has to be thermalized and
a neutrino that leaves the system, cooling the disk even more. Our model has
axial symmetry, the azimuthal velocity being keplerian, and hydrostatic
equilibrium holds in the direction perpendicular to the plane of the disk,
with the pressure given by the nucleons. The disk is heated by viscous
processes and locally cooled by radiation and by nuclear reactions. This
innermost region is surrounded by a region in which matter is completely
depleted of He, due to breakup. Out of this neighboring region, we assume
the existence of some unknown process that heats the gas and makes He
breakup viable. Since the nucleon temperature should be close to the virial
one for charge exchange reactions to occur, the disk will be geometrically
thick, and advection should be important. However, the more important
advection is, the less important is the radiative cooling, and the greater
the nucleon temperature will be, a fact that favors the onset of nuclear
reactions. Contradictory as it may seem, we don't know by sure the effect on
the nuclear reaction rate by the neglecting of advection. Besides this, 
\citet{esin97} have studied the behavior of advection as a function of the
accretion rate and of the viscosity parameter, concluding for its importance
of for small values of ${\dot{m}}$ (in Eddington units). As ${\dot{m}}$
increases, from $0.01$ to a critical value that depends on the viscosity
parameter, the advection zone becomes very luminous. The advection zone
shrinks in size with further increase of the accretion rate, and for
sufficiently large values it disappears, with the thin disk extending down
to the marginally stable orbit. In this paper we are interested in order of
magnitude estimates, and accounting of advection would make our calculations
very complex. \citet{mei91a} has taken conductive transport into account
in the inner geometrically thick region of a two temperature soft
Comptonized accretion disk in Cygnus X-1, showing that this process is much
more important than radiative cooling. For a matter of consistency, were
advection included, we should also have included conduction, which would
render the problem much more complicated.

\section{Disk Equations}

Hydrostatic equilibrium in the vertical direction reads 
$$
P={\frac{{\rho }}{3}}\,{\Omega }^{2}\,H^{2}\,\,,\eqno(3) 
$$%
where $\rho $ is the density, $\Omega $ is the Keplerian angular velocity
and $H$ is the semi-scale height of the disk. $P$, the total pressure, is
mainly given by the pressure due to the protons, i.e., 
$$
P=N\,10^{11}\,k\,T_{i}\,\,,\eqno(4) 
$$%
$N$ being the proton number density, $k$ the Boltzmann constant. Ion
temperature is in units of $10^{11}$ and electron temperature in units of $%
10^{9}$. Unless otherwise stated, physical variables are expressed in the
cgs system of units. Combining eq. (3) and (4) we obtain 
$$
H=4.93\times 10^{9}\,{\Omega }^{-1}\,{T_{i}}^{0.5}\,\,.\eqno(5) 
$$

Mass conservation, together with conservation of angular momentum and the
definition of the viscosity parameter yield 
$$
{\frac{{{\dot{M}}\,S}}{{4\,{\pi }}}}\,{\Omega }={\alpha }\,P\,H\,,\eqno(6) 
$$
which implies for the column density 
$$
{\Sigma }=1.96\times 10^{-20}\,{\alpha }^{-1}\,{T_i}^{-1}\,{\Omega }\,{\dot{%
M
}}\,S\,,\eqno(7) 
$$
$\alpha $ being the viscosity parameter, $\dot{M}$ is the accretion and $S$\
is a function that depends on the boundary condition imposed on the torque
at the inner radius. For $Q^{+}$, the heat flux, viscously generated, going
into the disk at a distance $R$ from the central object, 
$$
Q^{+}={3/{8\,{\pi }}}\,{\dot{M}}\,{\Omega }^2.\eqno(8) 
$$

In the two-temperature regime, energy is collisionally transferred from
protons to electrons and pairs at a rate, corrected for the inclusion of
neutrons, given by \citep{gui85,spi62} 
$$
{Q_{p}}^{-}=1.72\times 10^{28}\,{\rho }^{2}\,{\left( T_{i}-T_{e}\right) }\,{%
\left( 1+0.41\,{T_{e}}^{0.5}\right) }\,{\left( 1+2\,z+y_{+}\right) }\,{%
\left( 1+y_{n}\right) ^{-2}}{T_{e}}^{-3/2}\,H\,,\eqno(9) 
$$%
where $y_{n}$ is the total neutron abundance, $y_{+}$ is the abundance due
to pion production and $z$ is the positron density to proton density ratio.
The electrons and pairs, in turn, loose energy through unsaturated inverse
Comptonization of externally supplied soft photons. Like \citet{shap76}, 
we shall assume these photons are copiously supplied and that the
Kompaneetz $y$ parameter is constant and equal to $1$ \citep{ryb79}, i.e., 
$$
y={\left( 0.674T_{e}\right) }^{i}\,{{\tau }_{es}}^{j}=1,\eqno(10) 
$$%
where $i=1$ for $T_{e}{\leq {5.9}}$ and $2$ otherwise, $j=1$ for ${\tau }%
_{es}{\ \leq {1}}$ and $2$ otherwise. We shall assume that the intensity of
the soft photons coming from the external source overwhelms the intensity
due to photons internally produced (bremsstrahlung). With the inclusion of
neutrons, the electron scattering depth changes to 
$$
{\tau }_{es}={\frac{{\sigma }_{Th}}{{2\,m_{p}}}}\,{\Sigma }\,{\left(
1+2\,z+y_{+}\right) }\,{\left( 1+y_{n}\right) ^{-1}}\,\,,\eqno(11) 
$$%
${\sigma }_{Th}$ being the Thomson cross section for electron scattering and 
$m_{p}$ the proton mass. Clearly, this way of treating the radiative problem
in the disk relies heavily on the existence of an external soft photon
source, which we assume, from the very beginning, as granted. We should
emphasize, since our main concern is not the radiative problem, we adopt
this procedure for practical reasons: the spectrum of these soft photons,
after being upscattered in energy, through unsaturated inverse
comptonization in the inner region, reproduces fairly well the observed
spectrum of Cygnus X-1, in the 8-500 KeV region \citep{shap76}.

For the moment we postpone the discussion about the cooling of the disk due
to breakup and pion production. We next discuss pair production in the disk.
\section{Pair Production}

As we have emphasized before, our main concern in this paper is find out the
possibility of occurrence of nuclear reactions in the inner region of the
disk. We, therefore, construct a reasonable and realistic environment of
photons and pairs, in which the nuclear reactions take place. As we have
done in the case of radiative transport, we shall make some simplifications
when treating the pairs. We assume pairs are created mainly by Wien photons
interacting with Wien photons. However, instead of writing down the formal
pair equilibrium equation, we shall adopt a procedure similar to \citet{lian79} paper. 
This essentially consists in relating the photon chemical 
potential to the soft photon flux. This kind of procedure is valid as long
as the plasma is in Wien equilibrium \citep{sve84}, which implies ${\tau }%
_{{\gamma \gamma }}>1$. Below, we briefly review this procedure.

For $k\,T_{e}<m_{e}\,c^{2}$, $f_{\nu }$, the photon occupation number, obeys
the Kompaneetz equation. In that limit 
$$
f_{\nu }=24\,{e^{{\mu }_{\nu }}}\,x^{-s}\,{\left( 1+x+x^{2}/{2!}+x^{3}/{3!}%
+x^{4}/{4!}\right) }\,\,,\eqno(12) 
$$%
where ${\mu }_{\nu }$ is the photon chemical potential and $x={h\,{\nu }}/{\
k\,T_{e}}$. The occupation number $f_{+,-}$ for positrons and electrons may
be written, respectively,%
$$
f_{+}=e^{{\mu }_{+}}\,e^{-E_{+}},\,\,\eqno(13a) 
$$%
$$
f_{-}=e^{{\mu }_{-}}\,e^{-E_{-}}.\,\,\eqno(13b) 
$$%
Since the number of particles is conserved in the reaction 
$$
{\gamma }+{\gamma }\rightarrow e^{+}+e^{-}\,\,,\eqno(14) 
$$%
the chemical potential should satisfy 
$$
2\,{\mu }_{\gamma }={\mu }_{+}+{\mu }_{-}\,\,.\eqno{15} 
$$%
For Cygnus X-1 and related sources ${\it n}\approx 1$, which implies $s=4$.
The radiative cooling will be 
$$
F_{r}={\pi }\,\int_{{\nu }_{0}}^{3\,k\,T_{e}}\,{\frac{{2\,{\nu }^{3}\,f_{\nu
}}}{{c^{2}}}}d\,{\nu }\,\,.\eqno(16) 
$$

Now, using eqs. (12), (13), (14), (15) and (16) we obtain 
$$
N^2\,z\,{\left( 1+z\right) }={\left( \frac{{J\,c^2\,F_r}}{{{\left(
k\,T_e\right) ^4}\,276\,{\pi }}}\right) ^2}\,\,,\eqno(17) 
$$
where $J$ is given by 
$$
J=4\,{\pi }\,c\,{m_e}^2\,k\,T_e\,K_2{\left( \frac{{m_e\,c^2}}{{k\,T_e}}
\right) }\,\,,\eqno(18) 
$$
$K_2$ being the modified Bessel function of 2nd kind and we have substituted
in the slowly varying function $\ln {\left( {k\,T_e}/{h\,{\nu }_0 }\right) }$
for canonical values appropriate for Cygnus X-1.

It is worth mentioning that the radiative cooling now will be given by 
$$
F_{r}=Q^{+}-Q_{n} \, \, , \eqno(19) 
$$
where $Q_{n}$ is the nuclear cooling. Adopting the procedure as we did to
treat radiative cooling and pair production we have overestimated both
processes. Since the energy production is constant, this implies in
underestimation of the nuclear reaction rate.

\section{${}^4$He, pion and neutron production}

In the chain of reactions leading to the ${}^{4}He$ breakup 
\[
18.35\, MeV+p+{}^{4}He\rightarrow {}^{3}He+D\,\, 
\]%
\[
20.578\, MeV+p+{}^{4}He\rightarrow {}^{3}He+p+n\,\, 
\]%
\[
19.844\, MeV+p+{}^{4}He\rightarrow {}^{3}H+p+p\,\, 
\]%
\[
5.494\, MeV+p+{}^{3}He\rightarrow D+p+p\,\, 
\]%
\[
0.739\, MeV+p+{}^{3}H\rightarrow {}^{3}He+n\,\, 
\]%
\[
2.224\, MeV+p+D\rightarrow p+p+n\,\, 
\]%
the one that takes longer to occur is the first one. Once this reaction has
ignited, the other follow very rapidly. So, we assume the breakup dominated
by this reaction \citep{gue88} and use a prescription \citep{gou82,gou86,gue89,gue90} 
to write the neutron reaction rate for this reaction, 
$$
R_{b}=5.67\times 10^{-16}\,{T_{i}}^{-0.5}\,e^{-2.56/T_{i}}\,N^{2}\,\,.\eqno(20) 
$$%
This implies a neutron abundance, $y_{b}$, due to this reaction, given by, 
$$
y_{b}=1.05\times 10^{9}\,{T_{i}}^{-0.5}\,e^{-2.56/T_{i}}\,{\Sigma }%
^{2}\,R^{2}\,H^{-1}\,{\ \left( 1+y_{n}\right) }^{-1}\,\,.\eqno(21) 
$$%
The total neutron abundance $y_{n}$ will be given by 
$$
y_{n}=y_{+}+y_{b}\,\,.\eqno(22) 
$$%
Using eq. (20) we can express the nuclear cooling due to this reaction as 
$$
Q_{ei}=2.84\times 10^{20}\,{\left( 1+1.37\,T_{i}\right) }\,{T_{i}}^{-1}\,{%
\Sigma }^{2}\,{\left( 1+y_{n}\right) }^{-2}\,\,.\eqno(23) 
$$

\begin{figure}
\plotone{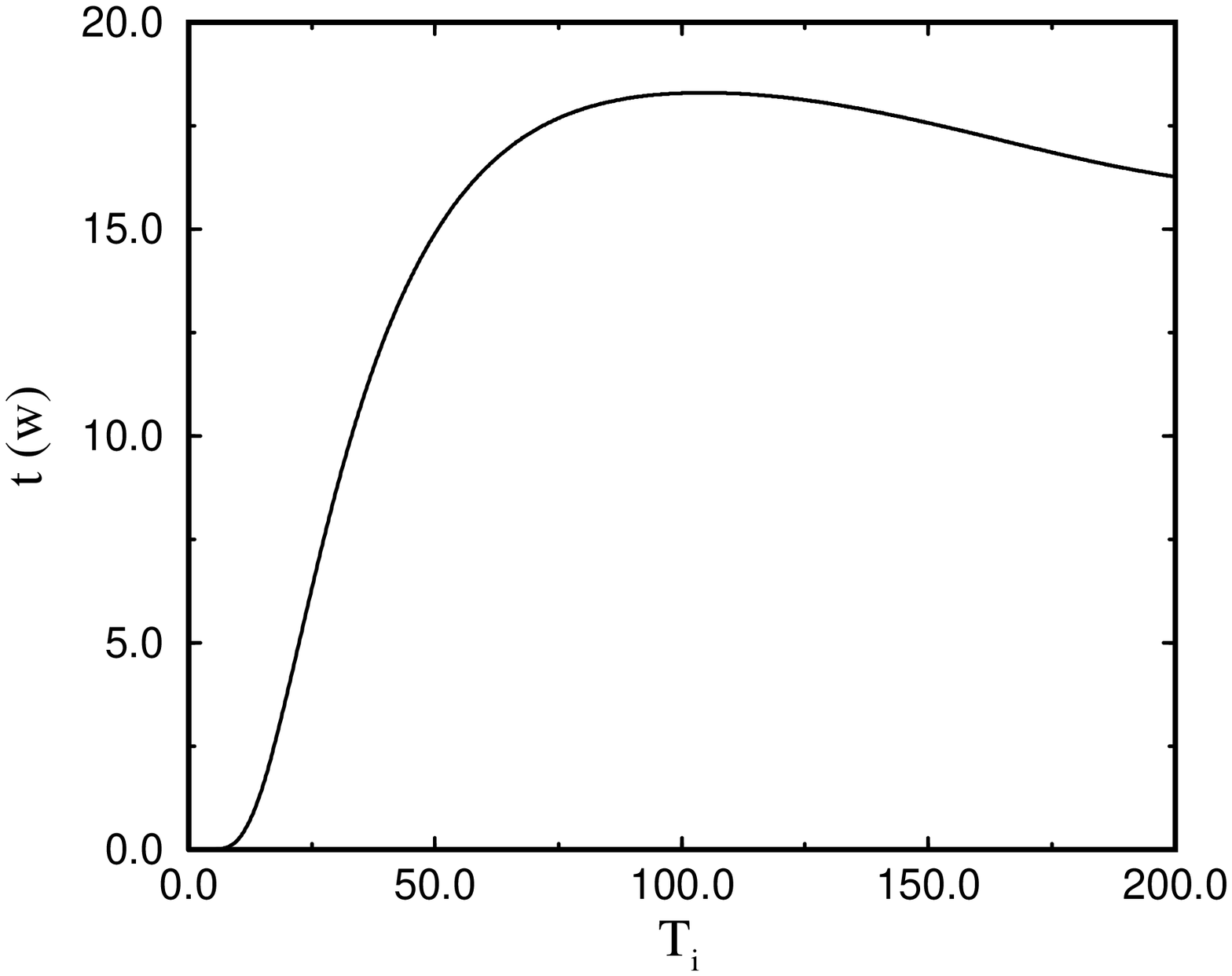}
\caption{$t(w)$, the reaction rate divided by the thermal speed of the
neutrons in units of {\em mb} \label{fig1}}
\end{figure}

For the reaction 
\[
p+p\rightarrow p+n+{\pi }^{+}\,\, 
\]%
we shall use\ the results by \citet{eng96}, who have 
carried out a very detailed study of the $p\,{\left( p,\,n{\pi }^{+}\right) }%
p$ and $p\,{\left( p,\,p{\pi }^{0}\right) }p$ reactions using a fully
relativistic Feynman diagram technique. Their calculations have been carried
out under the prescription of the one boson exchange (OBE) model, with
allowance for the inclusion of both nucleon and delta isobar excitations in
the intermediate states, as well as for the exchange of $\pi $, $\sigma $, $%
\rho $, and $\omega $ mesons. Accounting for the exchange of $\sigma $ and $%
\omega $ makes their results very reliable close to the threshold, where
these contributions presumedly dominate. Most of the parameters of the OBE
model are determined by fitting to the N-N scattering data over the energy
range of 300 MeV to 2 GeV. A fitting to their data for the total cross
section leads to%
$$
{\sigma}_{pp}=10^{f\left(E\right)}\,\,.\eqno(24)
$$
where 
$$
f\left( E\right) =\left\{ 
\begin{array}{lll}
0.0371785E-14.781764 & , & 290.0\leqslant E\leqslant {378.8} MeV \\ 
-4.96247+0.0148283E-0.00000839163E^{2} & , & \,378.8\leqslant E\leqslant {%
1060} MeV \\ 
-0.000064893617E+1.430787 & , & 1060\leqslant E\leqslant {2000} MeV,%
\end{array}%
\right. 
$$
where $E$ is expressed in MeV and the cross section in mb. The reaction rate
for this reaction will be written as 
$$
{R}_{+}=0.5\,N^{2}\,{\left( {\frac{{k\,T_{i}}}{m_{p}}}\right) }%
^{0.5}\,\int_{w_{s}}^{\infty }{\sigma }_{pp}\,e^{-w}\,w\,dw\,\,,\eqno(25) 
$$%
$w$ is the energy in units of $k\,T_{i}$ and $w_{s}$ is the threshold energy
for this reaction. Then, $y_{+}$, the contribution of this reaction to the
neutron abundance, follows, straightforwardly, 
$$
y_{+}=9.24\times 10^{-4}\,R^{2}\,{\Sigma }^{2}\,{\left( {\frac{{k\,T_{i}}}{%
m_{p}}}\right) }^{0.5}\,t{\left( w\right) }\,{\dot{M}}^{-1}\,H^{-1}\,{\left(
1+y_{n}\right) }^{-1}\,\,,\eqno(26) 
$$%
where $t{\left( w\right) }$ is the integral defined in eq. (25), now in
units of $10^{-27}$ mb. Figure \ref{fig1} shows $t\left( w\right) $ as a function of $%
T_{i.}$ 

To write down the cooling of the disk due to this reaction, we remind
ourselves that the pion has a mean life of about $2.6\times 10^{-8}\,$s,
decaying through 
\[
{\pi }^{+}\rightarrow {\mu }^{+}+{\nu }_{\mu }+34\, MeV\,.\, 
\]%
The ${\mu }^{+}$, in turn, has a mean life of about $2.2\times 10^{-6}$ $s$,
decaying through 
\[
{\mu }^{+}\rightarrow {\nu }_{e}+e^{+}+\bar{\nu}_{\mu }+105\, MeV\,,\, 
\]%
which allows us to write for the cooling%
$$
Q_{pp}=1.23\times 10^{29}\,{\left( 2.40\times 10^{-2}+2.07\times
10^{-7}\,T_{e}\right) }\,{\Sigma }^{2}\,{T_{i}}^{0.5}\,t\,{\left( w\right) }%
\,H^{-1}\,{\left( 1+y_{n}\right) }^{-2}\,\,.\eqno(27) 
$$

Finally, we can verify if neutrons can respond for the viscosity in the
disk. In order to do so, we write for the neutron viscosity \citep{wea76}
$$
{\eta }=1/3\,{\rho }_{n}\,v_{n}\,{\lambda }_{n}\,\,,\eqno(28) 
$$%
where ${\rho }_{n}$, $v_{n}$, and ${\lambda }_{n}$ are respectively neutron
density, neutron velocity and neutron mean free path. Taking the average for
a Maxwellian distribution, we obtain \citep{bond65}
$$
{\nu }_{n}=1.08\,y_{n}\,{\left( \frac{{k\,T_{i}}}{m{n}}\right) }^{.5}\,{\ell 
}_{n}\,\,,\eqno(29) 
$$%
${\ell }_{n}$ being the effective mean free path given by 
$$
{\ell }_{n}={\frac{{\lambda }_{n}}{{1+{{\lambda }_{n}}^{2}}}}\,\,,\eqno(30) 
$$%
with ${\lambda }_{n}$ given by 
$$
{\lambda }_{n}={\frac{m_{p}}{{{\sigma }_{n}\,{\rho }\,{\left( 1+y_{n}\right) 
}}}}\,\,.\eqno(31) 
$$%
The cross section for neutron scattering, averaged over a Maxwellian
distribution, is \citep{gam63}
$$
{\sigma }_{n}=5.47\,{T_{i}}^{-0.85}\,b\,\,.\eqno(32) 
$$%
Finally, using the definition of the viscosity parameter $\alpha $ together
with eq. (29) we arrive at 
$$
{\alpha }^{2}=3.65\times 10^{-2}\,{\dot{M}}_{17}\,y_{n}\,{\Omega }\,{T_{i}}%
^{-1.85}\,{\left( 1+y_{n}\right) }^{-1}-10^{-5}\,{{\dot{M}}_{17}}^{2}\,{%
\Omega }^{2}\,{T_{i}}^{-3.7}\,\,,\eqno(33) 
$$%
the accretion rate expressed in units of $10^{17}\,$g\thinspace s$^{-1}$.

\begin{figure}
\plotone{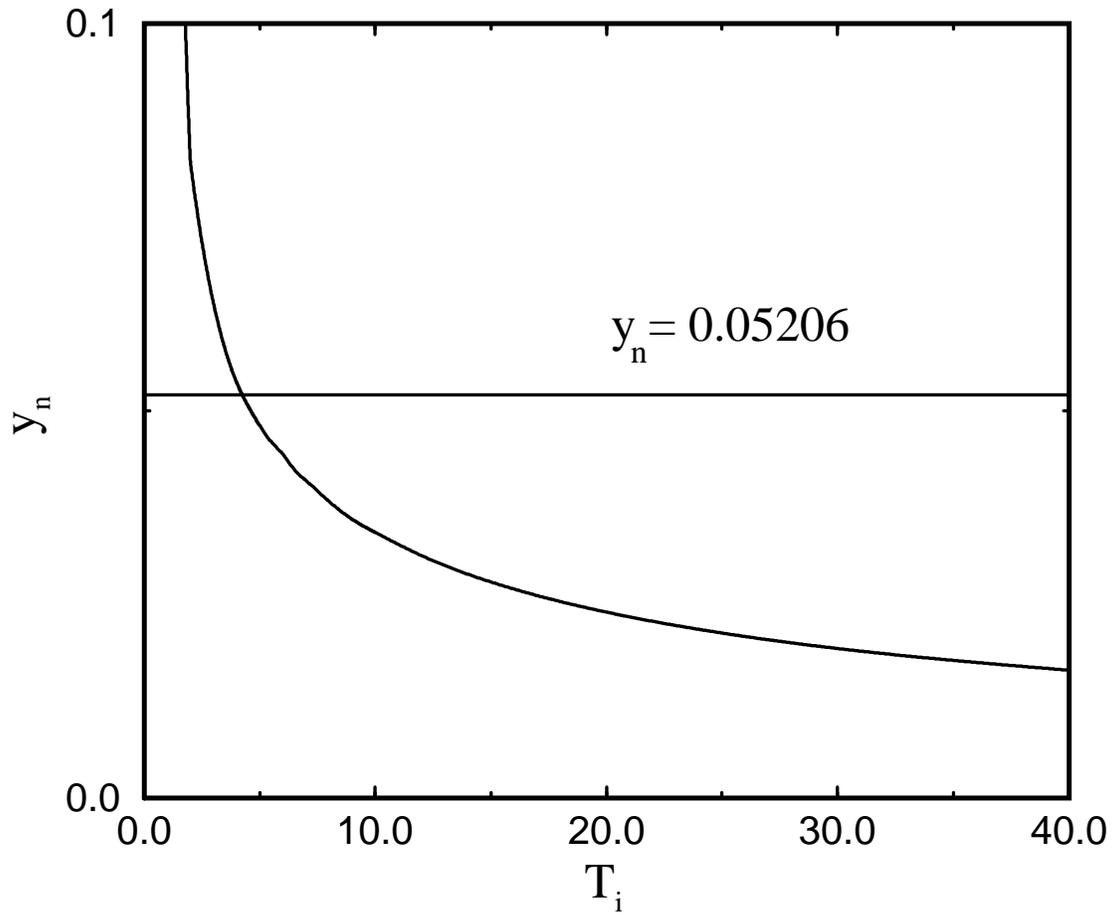}
\caption{The neutron abundance $y_{n}$, as a function of the ion
temperature, calculated at $r=10$ for ${\dot{M}}_{17}=1.0$. For $T_{i}=4.16$
there is total ${}^{4}$He depletion. \label{fig2}}
\end{figure}

We are now in a position to ask ourselves if neutrons collisions with the
accreting matter can indeed act as a source for the viscosity in the
innermost regions of an accretion disk. We shall see that previous results
\citep{mei93} relied heavily on the constraint imposed assuming equality
between nuclear reaction time and dynamical time. Dropping this assumption
leaves practically no restriction as far as the accretion rate is concerned.
To not be burdened by additional complexities let us unravel things a little
bit by neglecting, for the moment, the contribution to the neutrons due to
pion production. Under this procedure, combination of eqs. (5), (7), (21)
and (33) yields 
\[
{y_{n}}^{2}\,{\left( 3.65\times 10^{-2}\,{\dot{M}}_{17}\,{\Omega }\,{T_{i}}%
^{1.15}-10^{-5}\,{{\dot{M}}_{17}}^{2}\,{\Omega }^{2}\,{T_{i}}^{-0.7}\right) }%
\,\, 
\]%
$$
-y_{n}\,10^{-5}\,{{\dot{M}}_{17}}^{2}\,{\Omega }^{2}\,{T_{i}}%
^{-0.7}-0.458\,e^{-2.56/T_{i}}=0\,\,.\eqno(34) 
$$%
Now specializing for $r=10$, we obtain 
$$
{\dot{M}}_{17}{\leq {5.77\,{T_{i}}^{1.85}}}\,\,.\eqno(35) 
$$%
This, however, is not the final result since we have to investigate the
effects of cooling and pair production. Figure \ref{fig2} shows the neutron abundance 
$y_{n}$ as a function of $T_{i}$ for $r=10$ and $\dot{M_{17}}=1.0$. 

\begin{figure}
\plotone{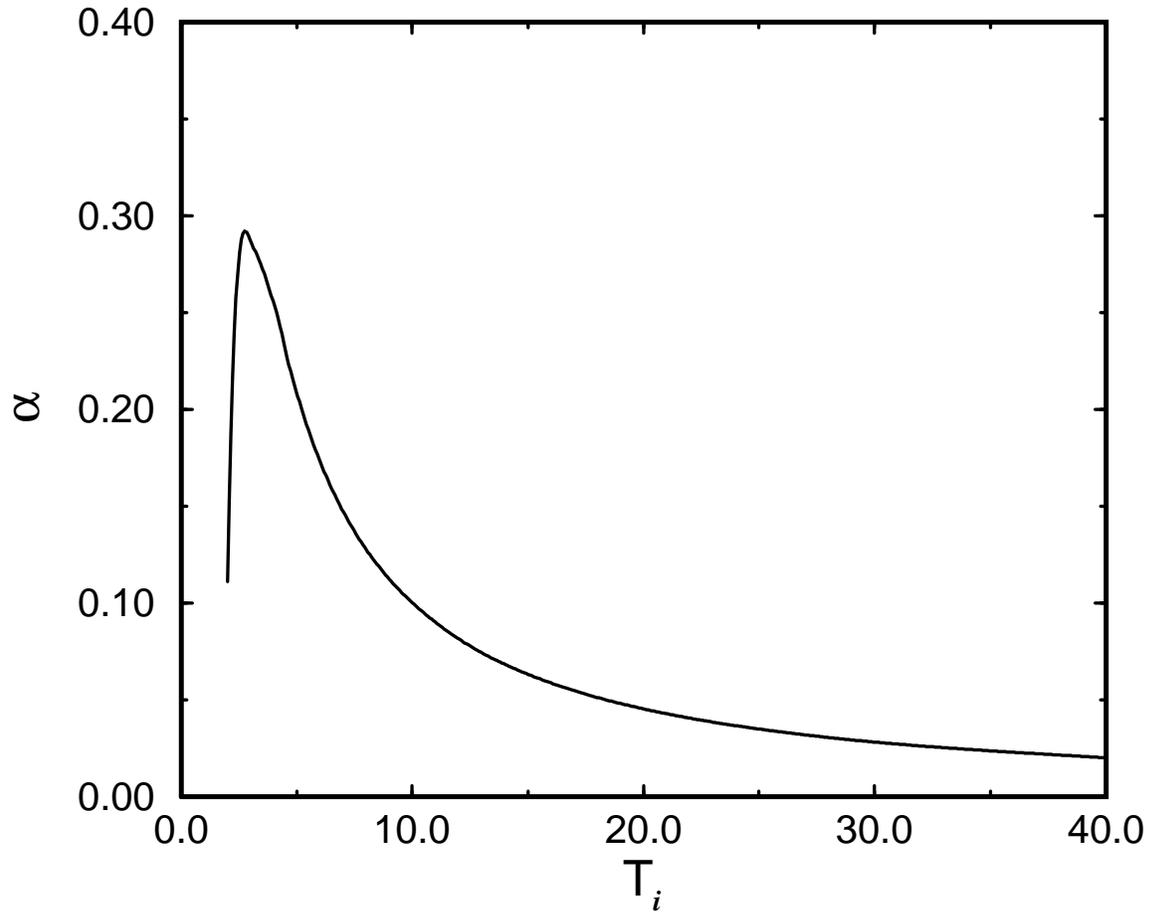}
\caption{The viscosity parameter $\protect\alpha $, as a function of the
temperature, calculated at $r=10$ for ${\dot{M}}_{17}=1.0$. \label{fig3}}
\end{figure}

In Fig. \ref{fig3} we present the viscosity parameter $\alpha $ as a function of $T_i$
in the same conditions of Fig. \ref{fig2}. We see that the viscosity parameter may
reach very high values. The maximum $\alpha $ occurs for $T_i\approx 3.0$. 

We finally reduce our system of equations for the disk to only two equations
involving $T_{e}$ and $T_{i}$. The first of them is the thermal equilibrium
equation and reads 
\[
{\left( 1+1.37{T_{i}}\right) }\,e^{-2.56/T_{i}}\,{\alpha }^{-2}\,{T_{i}}%
^{-3}\,{\left( 1+y_{n}\right) }^{-2}\,\, 
\]%
$$
+12.14\,{\left( 1+0.41\,{T_{e}}^{0.5}\right) }\,{\alpha }^{-1}\,{T_{i}}%
^{_{0}.5}\,{T_{e}}^{2.5}\,{\left( 1+y_{n}\right) }^{-1}-10.92=0\,\,,\eqno%
(36) 
$$%
and the second is the pair equilibrium equation, 
$$
{\Delta }^{2}-1-252.5\,{T_{i}}^{2}\,{\left( 1+0.41{T_{e}}^{0.5}\right) }%
^{2}\,e^{-11.87/T_{e}}\,{T_{e}}^{-10}=0\,\,,\eqno(37a) 
$$%
where $\Delta $ is given by 
$$
{\Delta }=6.26\,{\alpha }\,T_{i}\,{T_{e}}^{-1}\,{\left( 1+y_{n}\right) }%
^{1}-y_{n}\,\,,\eqno(37b).
$$%
${\alpha }$ and $y_{n}$ are given respectively by eqs. (33) and (34).

\begin{figure}
\plotone{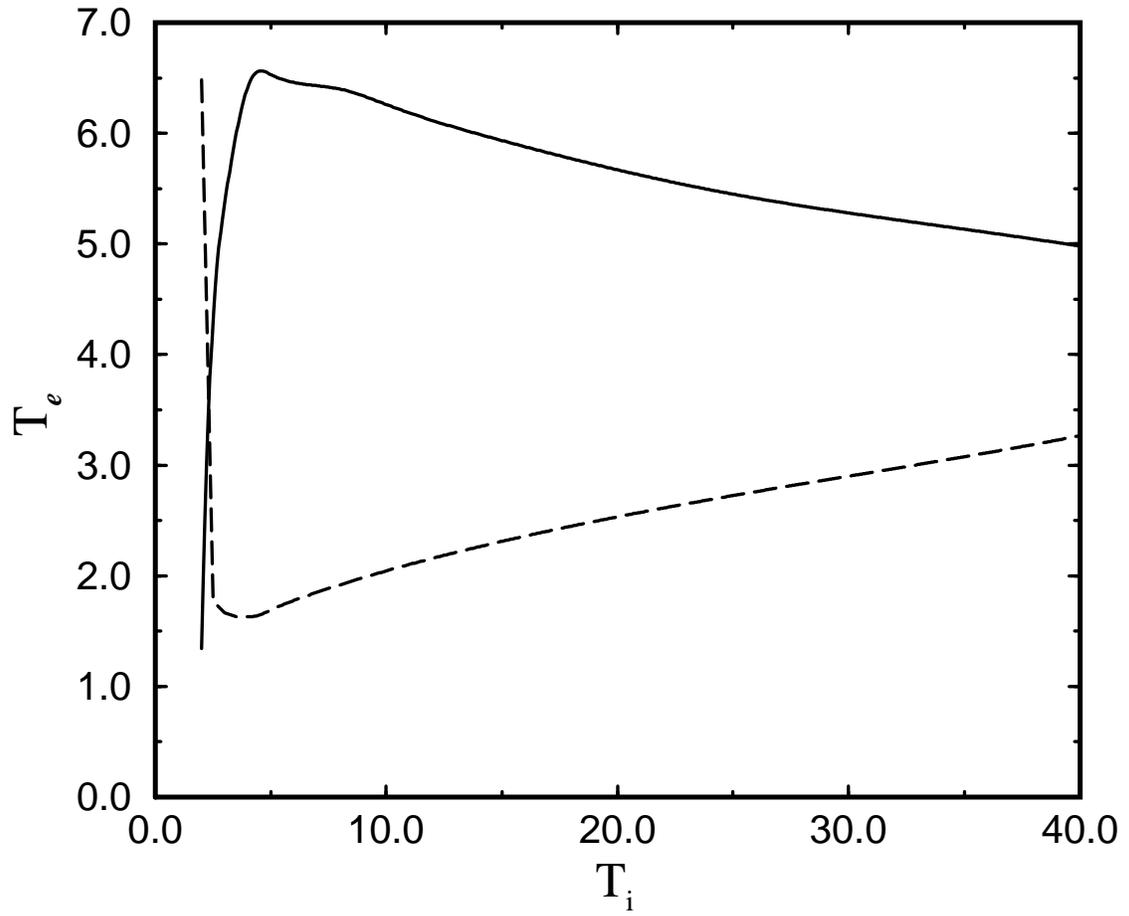}
\caption{Relation $T_e-T_i$ obtained through thermal equilibrium (lower
curve) and through pair equilibrium (upper curve). \label{fig4}}
\end{figure}

A glance at Fig. \ref{fig4} reveals the existence of a solution for the
two-temperature soft photon Comptonized accretion disk with pairs and
viscosity generated by neutron collisions with the accreting matter,
neutrons supplied by ${}^4$He breakup. 

\begin{figure}
\plotone{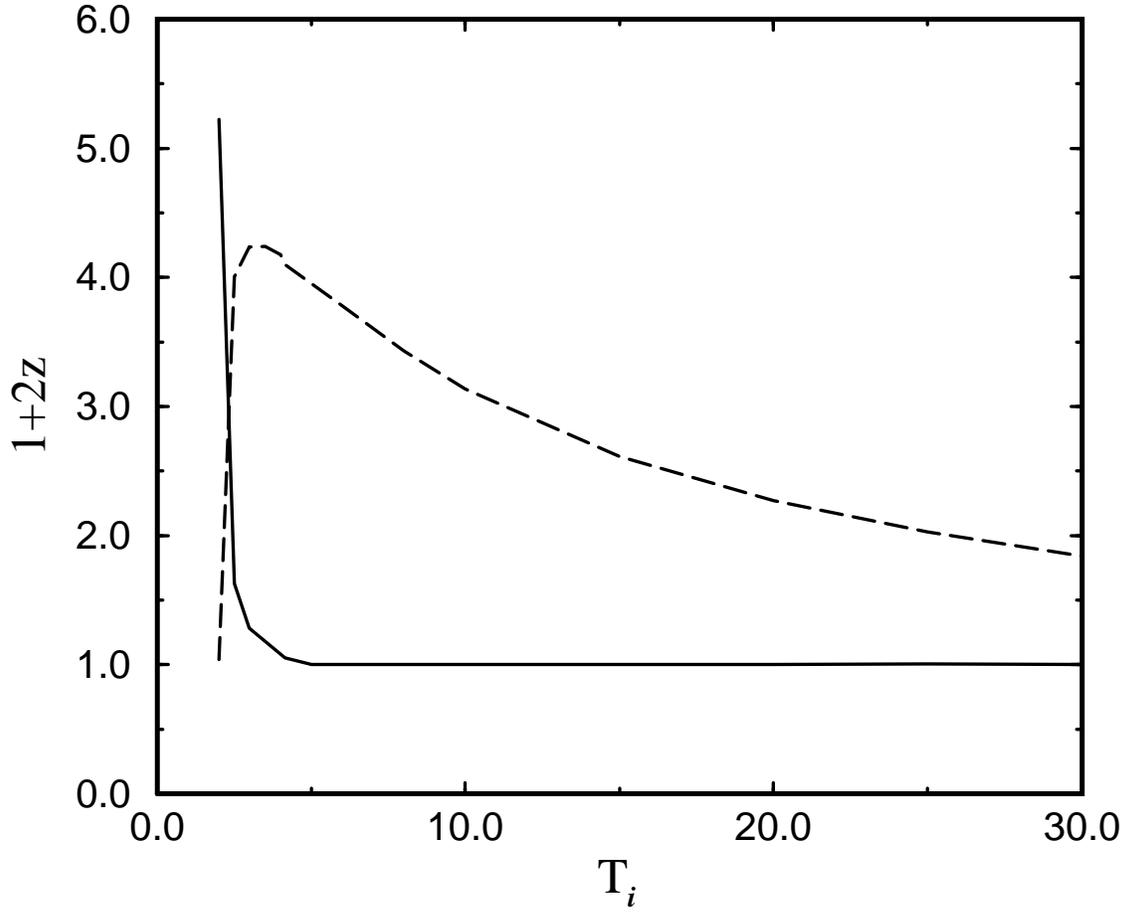}
\caption{Relation between $(1+2\,z)$ and $T_i,$ for $\dot{M}_{17}=1.0$. The
upper curve is obtained using $T_i-T_e$ data from the solution of the
thermal equilibrium equation; the lower, using data from the pair
equilibrium equation. \label{fig5}}
\end{figure}

To check for the consistency of our solution we also plot $(1+2z)$ in Fig.
\ref{fig5}, both in the thermal equilibrium and pair equilibrium situations. 

\section{Highlighting pion production}

We have seen in the previous section, neglecting contribution from pion
production to the neutrons in the disk, the existence of a solution to the
disk equations at $r=10$, $T_{i}\approx 2.5$. Were pion production included,
the results would not change so much, since at this temperature the
contribution would be negligible. However, at that temperature, the
depletion of ${}^{4}$He is complete. Therefore, for $r<10$, breakup no
longer contributes to the cooling of the disk, its contribution being
restricted to the viscosity through the neutrons already produced.

To underline the pion contribution relative to the neutron production let us
calculate it at the point where the ion temperature is largest, i.e., $r=1$.
Calculating physical variables at this point and inserting in eq. (26)
yields 
$$
{y_{+}}^{2}+{20/19}\,y_{+}-0.364\,{\alpha }^{-2}\,{T_{i}}^{-2}\,t{\left(
w\right) }\,\,,\eqno(38) 
$$%
and ${\alpha }$ is now (eq. (33)) 
$$
{\alpha }^{2}=730\,{\left( {\frac{{1+19\,y_{+}}}{{20+19\,y_{+}}}}\right) }\,{%
T_{i}}^{-1.85}-4\times 10^{3}\,{T_{i}}^{-3.7}\,\,.\eqno(39) 
$$

\begin{figure}
\plotone{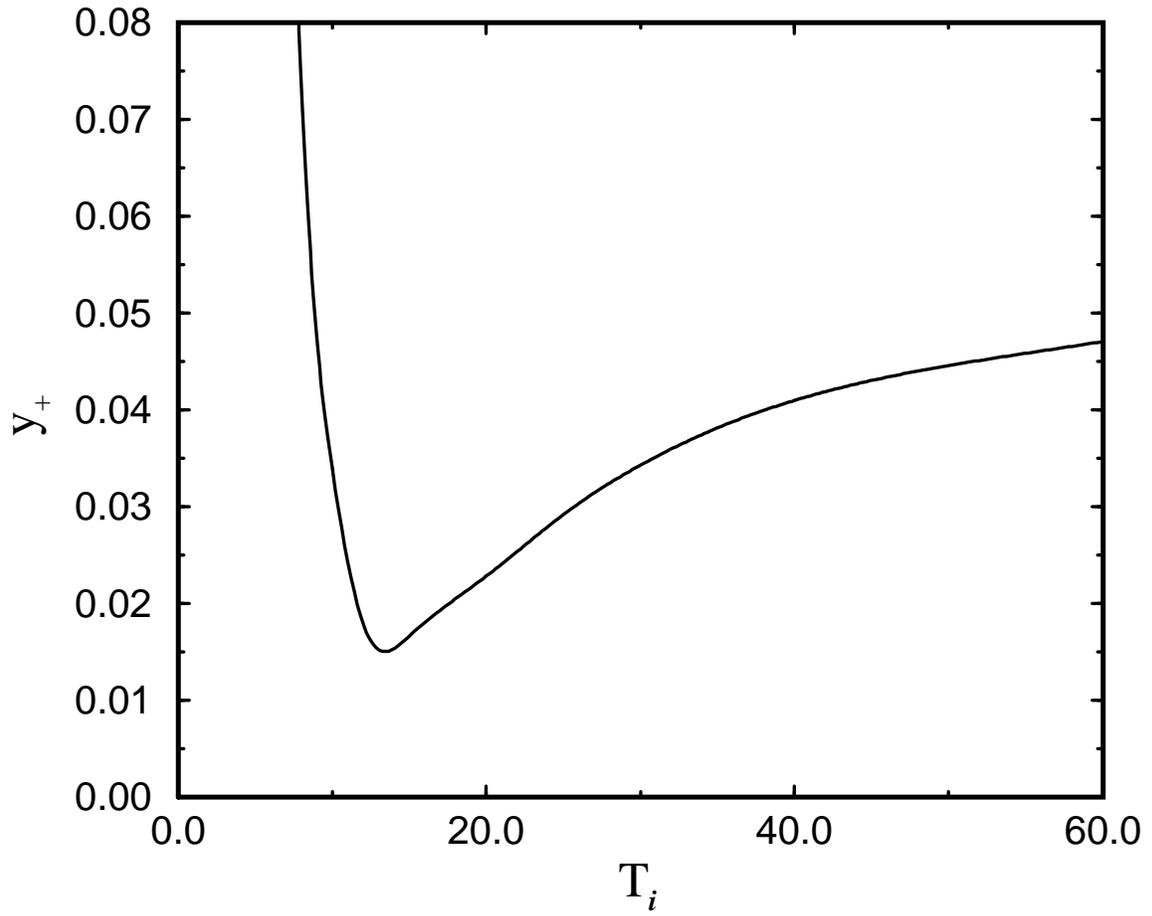}
\caption{$y_{+}$, the $p\,{\left( p\,{\protect\pi }^{+}\,n\right) }\,p$
contribution to the neutron abundance. \label{fig6}}
\end{figure}

\begin{figure}
\plotone{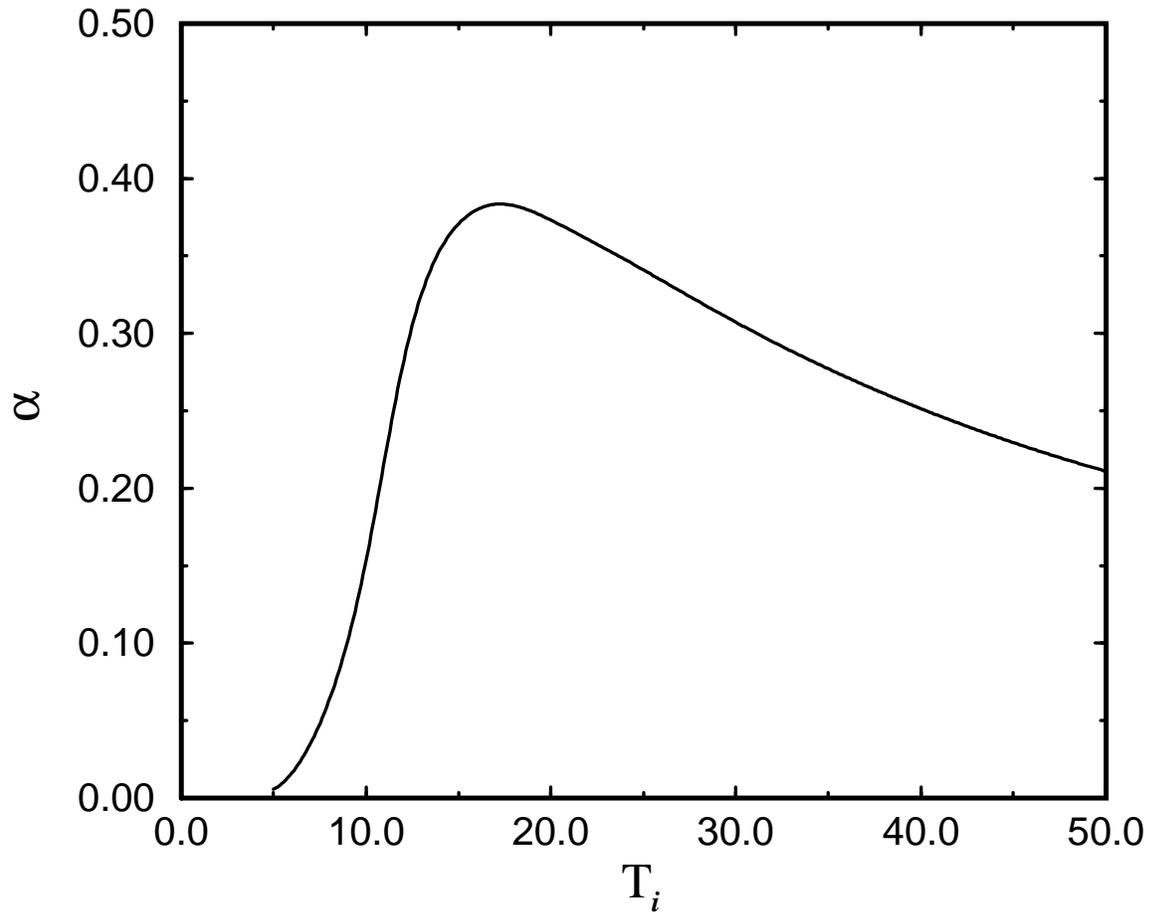}
\caption{The viscosity parameter\ ${\protect\alpha }$ taking{\bf \ }pion
production into account at $r=1$. \label{fig7}}
\end{figure}

To see how ${\alpha }$ and $y_{+}$ behave in that region, we plot them in
Figs. \ref{fig6} and \ref{fig7}. 

In Fig. \ref{fig6} we see that production of neutrons due to $p\,{\left( p\,{\pi }%
^{+}\,n\right) }\,p$ may be comparable to the contribution due to the
breakup. 

Looking at eq. (35) we realize that the constraint on the accretion rate now
reads 
$$
{\dot{M}}_{17}{\le {0.183\,{T_i}^{1.85}}}\,\,.\eqno(40) 
$$

As we have done previously, we now investigate the effects of cooling and
pair production and, for that, we reduce our system of equations to only two
equations involving $T_e$ and $T_i$. The thermal equilibrium and pair
equilibrium equations turn to, respectively 
\begin{eqnarray}
&&{\left( 20/19+y_{+}\right) }^2-3.86\,{\alpha }^{-2}\,{T_i}^{-2}\,t{\left(
w\right) }-  \nonumber \\
&&1.1\,{\alpha }^{-1}\,{T_i}^{-0.5}\,{T_e}^{-2.5}\,\,{\left( 1+0.41{T_e}%
^{0.5}\right) }\,{\left( 20/19+y_{+}\right) }=0  \nonumber
\end{eqnarray}
and 
$$
{\Delta }^2-1-204\,{T_i}^2\,{\left( 1+0.41{T_e}^{0.5}\right) }%
^2\,e^{-11.87/T_e}\,{T_e}^{-10}=0\,\,,\eqno(41) 
$$
where $\Delta $ is now given by 
$$
{\Delta }=0.1975\,{\alpha }\,T_i\,{T_e}^{-1}\,{\left( 20/19+y_{+}\right) }%
-y_{+}\,\,,\eqno(42) 
$$
and $y_{+}$ and ${\alpha }$ are given respectively by eqs. (38) and (39).

\begin{figure}
\plotone{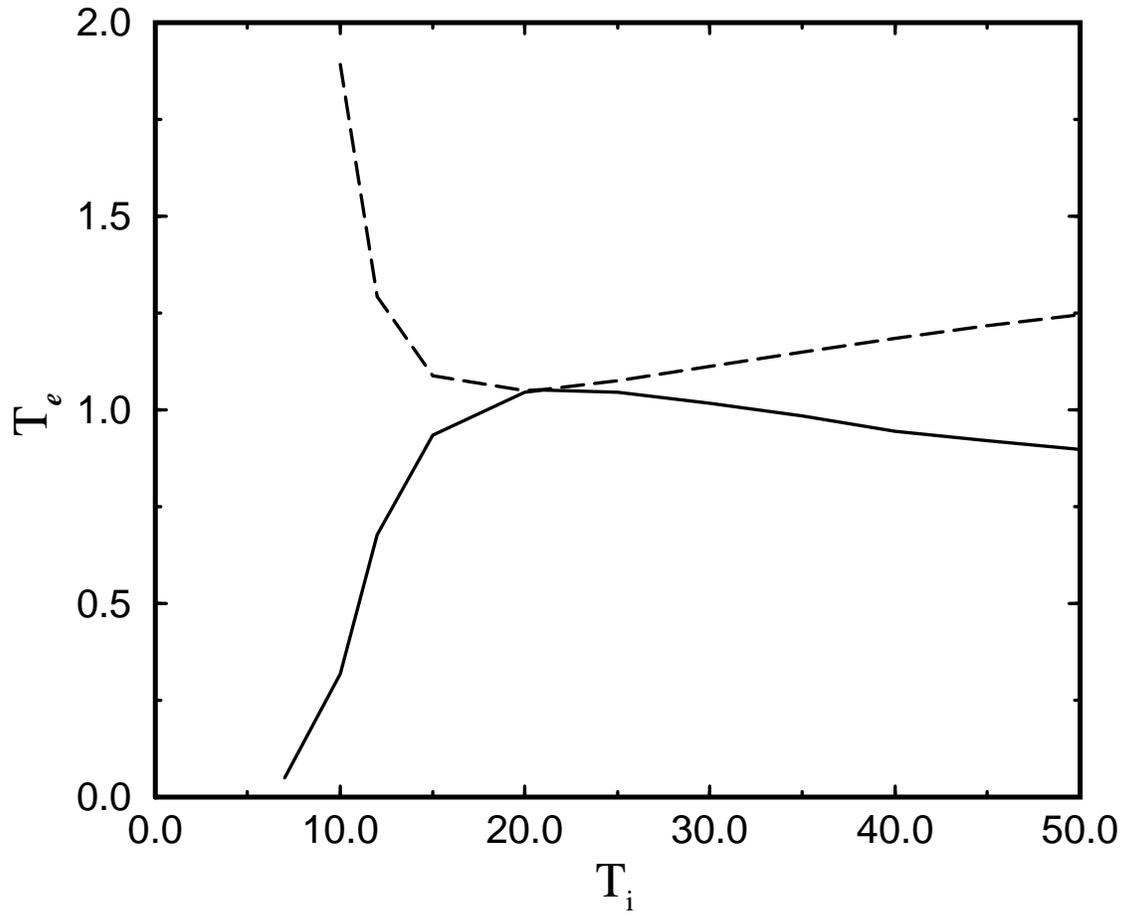}
\caption{ The solution to the thermal equation (upper curve) and the pair
equilibrium equation (lower curve). The curves are tangent at $T_i\approx
21. $ \label{fig8}}
\end{figure}

In Fig. \ref{fig8} we plot the solutions to the thermal and pair equilibrium
equations. 

\begin{figure}
\plotone{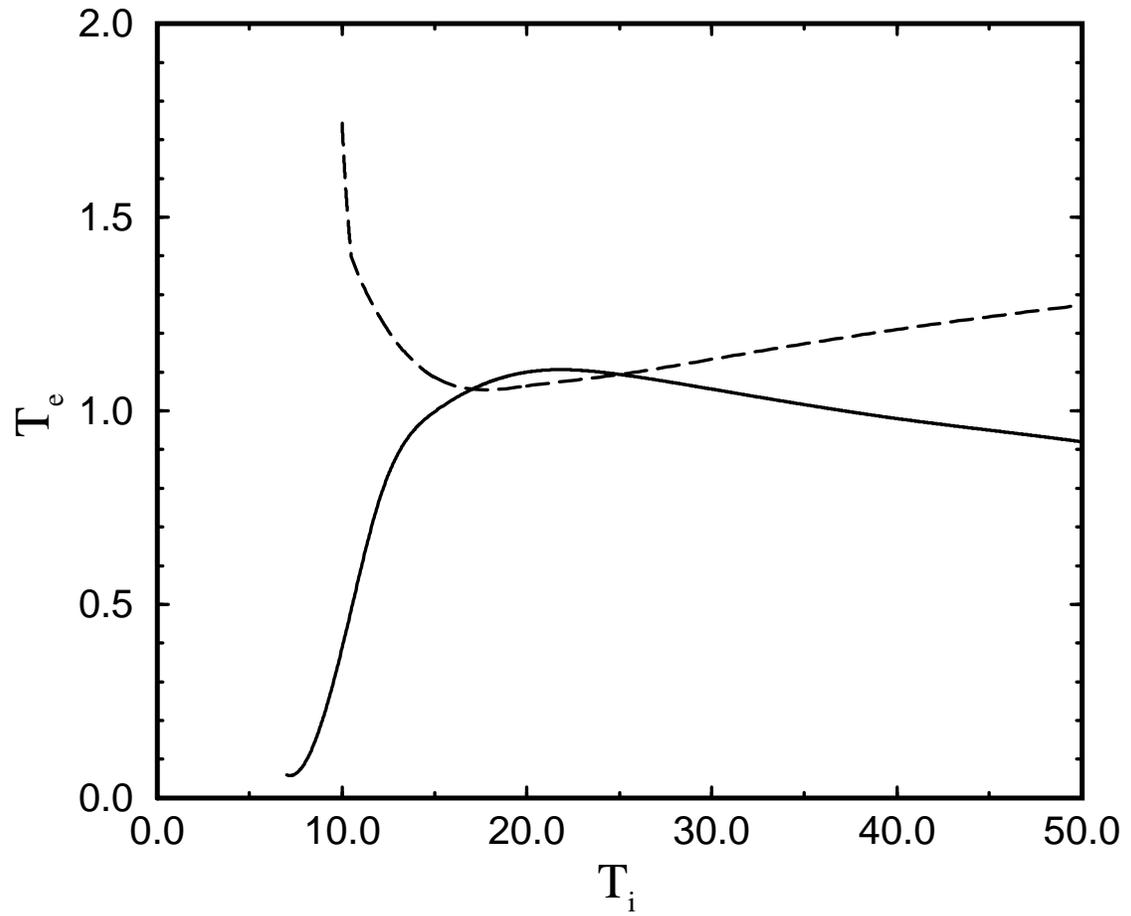}
\caption{The solutions to the thermal and pair equilibrium equations for ${%
\dot{M}}_{17}=0.9$. Contrary to the previous figure, there are two
solutions. \label{fig9}}
\end{figure}

To see how the solutions of these equations behave as a function of the
accretion rate, we have plotted them for ${\dot{M}}_{17}=0.9$ in Fig. \ref{fig9}.
There it can be seen the emergence of two solutions instead of only one as
in the ${\dot{M}}_{17}=1.0$ case.

\begin{figure}
\plotone{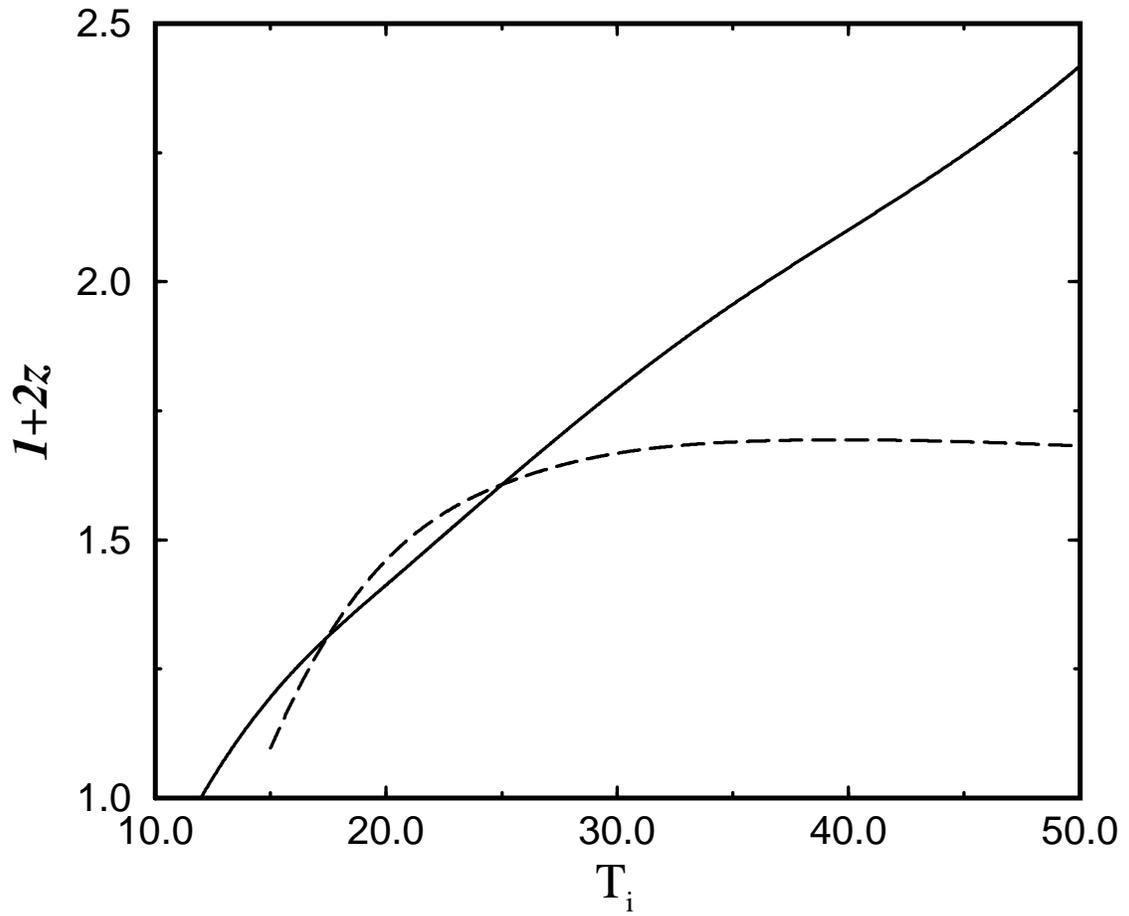}
\caption{$(1+2\,z)$ for ${\dot{M}}_{17}=0.9$. The curve with larger values
at the right is obtained using the data from the pair equation, the other
using data from the thermal equation. \label{fig10}}
\end{figure}

In Fig. \ref{fig10} we plot ${(}1+2\,z)$ as a function of $T_i$. In contrast to Fig.
\ref{fig5}, we have now two solutions, both with $z<1.0.$ 

\begin{figure}
\plotone{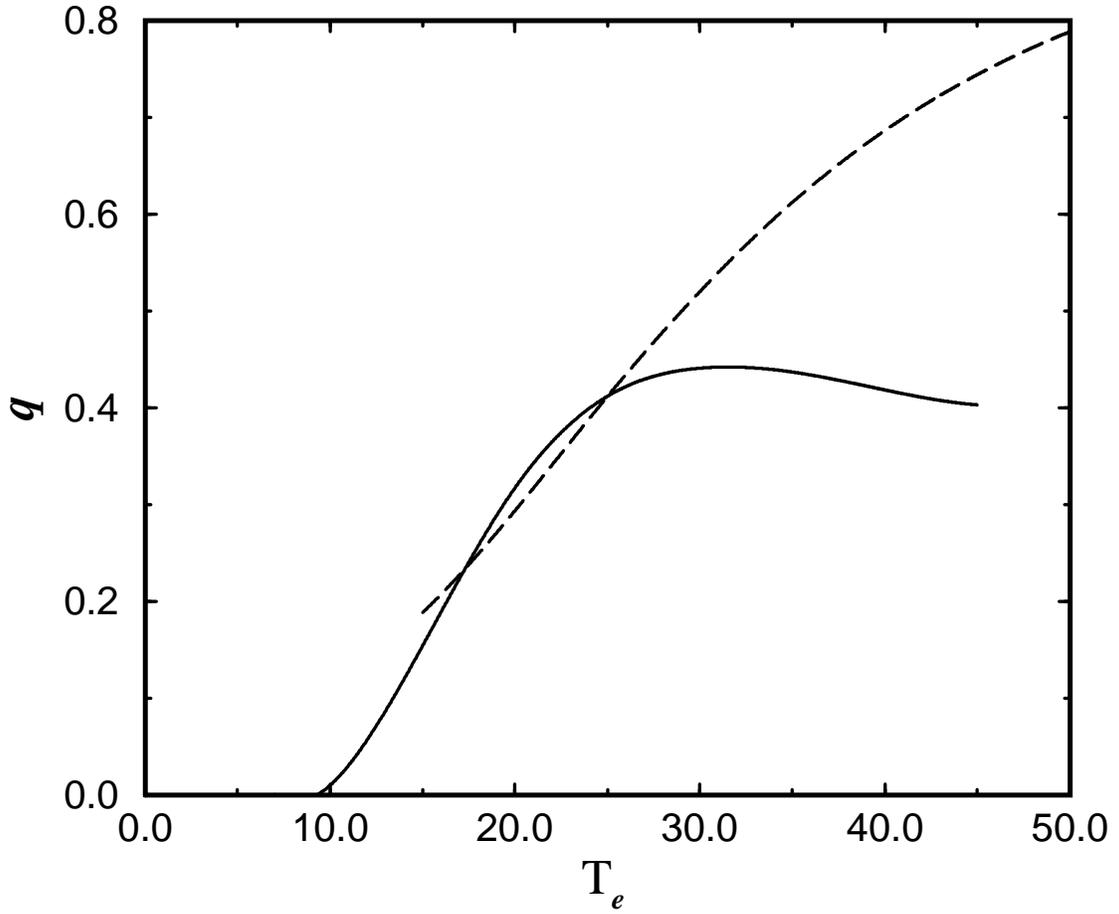}
\caption{$q=Q_{pp}/Q_{ei}$ for ${\dot{M}}_{17}=0.9$. The upper curve on the
right is built with data from thermal equation, the other with data from the
pair equation. For $T_i<12$ the curve coming from the thermal data grows
steeply (not shown here). \label{fig11}}
\end{figure}

Finally, to see the importance of nuclear cooling due to ${\pi }^{+}$
production, we plot $q=Q_{pp}/Q_{ei}$ in Fig. \ref{fig11}. 

\section{Analysis and conclusions}

We don't know by sure what physical processes may be operating in the
outermost parts of the disk generating viscosity and switching accretion on.
Putting this flaw aside, we realize that under certain conditions protons
and electrons are out of thermal equilibrium, the ion temperature being much
greater and close to the virial one and, for a matter of self-consistency,
we are compelled to consider some nuclear reactions, since we have available
energy up to twice the proton rest mass. However, besides energy
considerations, we had a very strong motivation to consider $^{4}$He breakup
and $p\,{\left( p\,n\,{\pi }^{+}\right) }\,p$. This is related to the
viscosity problem itself: these reactions produce neutrons whose collisions
with the accreting matter may be a source of the required viscosity to drive
accretion. We have considered these reactions in an environment of protons,
electrons, neutrons, photons and pairs. Concerning radiative cooling and
pair production, we have made some simplifications that overestimated these
processes and, as a consequence, underestimated the nuclear reaction rate
because the total energy available is constant. The treatment we have
adopted to calculate the nuclear reaction rate consists of an improvement
over a previous one \citep{mei93} due to the abandon of the assumption of
equality between nuclear reaction and dynamical times, as well as for the
inclusion of both pion and pair production. To emphasize the role of the
nuclear reactions, as far as the requirement of huge ion temperatures needed
to ignite pion production, we have assumed a maximally synchronously
rotating Kerr black hole, the inner radius of the disk equal to its horizon, 
$R_{h}$. The region of the disk we were concerned is the one extending from
the horizon to about $20\,R_{h}$. In the outer parts of the inner region,
the contribution of the reaction $p{\left( p,n{\pi }^{+}\right) }\,p$ to the
production of neutrons is negligible, this being dominated by${}^{4}$He
breakup, which we use as a kind of boundary condition. Neutrons produced by
this reaction contribute to the viscosity and the plasma heats up as we
approach the hole, making possible the production of pions which, in turn,
increase the viscosity by concomitant production of neutrons.

The radiative cooling of the disk was assumed to be due to unsaturated
inverse Comptonization of soft external photons impinging upon the disk.
Since we have not considered the reaction $p{\left( p,p{\pi }^{0}\right) }p$%
, we have not taken into account the radiative cooling due to photons coming
from the ${\pi }^{0}$ decay. Part of those soft photons are upscattered in
energy, reaching the Wien peak where they interact with themselves producing
pairs. Radiative cooling and pair production decrease the electron
temperature, an effect that hinders the efficiency of nuclear reactions,
even if the ion temperature is great enough for them to occur. When $T_{e}$
decreases, the number of interacting electrons with one proton increases
sharply, due to less shielding. Nuclear forces only act at short distances
and the high ion temperature needed to overcome the Coulomb barrier also
hinders the occurrence of nuclear events due to very short time in the
nuclear range. As a matter of fact, reaction rate due to electromagnetic
interactions prevails over the rate due to nuclear interactions if $T_{e}$
satisfies 
$$
T_{e}<28.56\,{T_{i}}^{1/3}\,\,.\eqno(43) 
$$%
However, the energy transferred in the (electromagnetic) scattering will be
in the KeV range, while for the nuclear reaction it will be in the MeV
range. Therefore, the previous inequality changes roughly to 
$$
T_{e}<0.2856\,{T_{i}}^{1/3}\,\,,\eqno(44) 
$$%
when one also considers the energy transferred in the event.

Aiming at the application to Cygnus X-1 and related sources, we have set ${%
\dot{M}}_{17}=1$ and ${\dot{M}}_{17}=0.9$ which are much less than the
critical value for ${\dot{M}}_{17}$ calculated close to the horizon, i.e., 
\[
{{\dot{M}}_{17}}^c=0.183\,{T_i}^{1.85}\,.\, 
\]
For both values of the accretion rate $T_e$ is larger than the value given
by the ineq. (43). As a result, for every solution we have found, $z<1$.
This is surprising, since it is known that, in the absence of these
reactions, one of the solutions for the two-temperature disk has $z>1$.
Another surprising result we have found concerns the sensitivity of the
number of solutions to the value of the accretion rate. Calculating at $r=1$
and $r=10$, for ${\dot{M}}_{17}=1$ we have found the solution is unique.
Decreasing to ${\dot{M}}_{17}=0.9$ there are two solutions. For these values
of the accretion rate the neutron abundance in the inner region is fairly
high, going from $y_n=1/19$ in the outer border, with total ${}^4$He
depletion, to about twice that value close to the horizon. This implies $%
\alpha $ quite large, i.e., large viscosity. This large viscosity,
nevertheless, comes along with a high nuclear cooling, being comparable to
the radiative cooling.

It is known that the innermost region of the accretion disk is secular and
thermally unstable \citep{shak76}in the absence of nuclear reactions. Taking into account
these nuclear reactions, the disk behaves in the same manner at the very
onset of these instabilities. However, it is worthwhile observing the
behavior of the viscosity parameter with $T_i$, because two scenarios may
emerge as far as the time evolution of the inner region is concerned. ${%
\alpha }$ starts growing, reaches a maximum somewhere between $T_i=15$ and $%
T_i=20$, then decreases and may become very small, even null, with
increasing $T_i$. At that moment, accretion starts switching off and the
cooling, that goes with ${\alpha }^{-2}$, is practically instantaneous.
Outside this region, however, some other mechanism for viscosity generation
is still operating and accretion, there, keeps going on. As the matter
reaches the border of the inner region it gets piled up there. As this
piling up goes on, this region will be subject to several instabilities.
Once one of these instabilities starts to grow it triggers accretion in the
inner region. As the accretion proceeds, the accretion rate decreases till
it reaches a value for which there are two steady states for the disk. Which
one will the disk choose? At that moment, do both thermal and pair
equilibrium hold in that region? A less drastic scenario is the one somehow
similar to that proposed by \citet{kus91}. Physical
processes in the disk are characterized by time scales, the one chosen by
the system being the least scale. As we have seen, thermal and pair
equilibrium only hold under special circumstances. It may happen that due to
the nuclear reactions and instabilities in the disk, the system may undergo
a kind of limit cycle behavior around the upper solution.

However, to have a better understanding of the time evolution of these
systems we should make better treatment of both radiative cooling and pair
production and a more detailed stability analysis of the problem as well.
That is what we intend to do next, in a future contribution, taking also
into account the reaction $p\,{\left( p\,\,p\,{\pi }^{0}\right) }\,p$.



\acknowledgments

C.L. Lima acknowledges support from FAPESP (project nr. 1998/065902)







\clearpage


\clearpage 

\clearpage

\clearpage

\clearpage

\clearpage

\clearpage

\clearpage

\clearpage

\clearpage

\clearpage

\clearpage


\end{document}